\documentclass{article}
\begin{document}
\baselineskip7mm
\title{Power-law anisotropic cosmological solution 
in 5+1 dimensional Gauss-Bonnet gravity}
\author {A.Toporensky and P.Tretyakov}
\date{}
\maketitle
\hspace{8mm} {\em Sternberg Astronomical Institute,
Universitetsky prospekt, 13, Moscow 119992, Russia}

\begin{abstract}
We write down an anisotropic solution for a flat 5+1 dimensional
Universe in Gauss-Bonnet gravity.  In the model under investigation
this solution replaces the generalized Kasner solution 
near a cosmological singularity.
\end{abstract}

It is known for more than 20 years that the famous Mixmaster chaotic behavior
near initial singularity \cite{BKL} is absent in space dimensions 10 and more \cite{henneaux}.
Later deep connections of this result with theory of infinite-dimensional Lie algebras
have been revealed \cite{henneaux2}. There is also an interesting fact (first noticed
in \cite{henneaux}) that the critical number of
space-time dimensions for this problem coincides with the critical dimensionality of 
superstring theory, though 
it is still unknown whether it is a pure coincidence or some profound reasons connecting these two problems
exist. An argument against a connections between these two properties of 10-dimensional gravity
is that the abovemensioned results on chaos have been established in Einstein gravity, though string gravity
corrections to Einstein equations (see, for example, \cite{Tseytlin, BentoB})
could change the dynamical behavior even in the framework of perturbation theory. There are several
papers on chaos in string gravity, and the results are different, depending on matter content
of the model \cite{Dambr, Damour, Rendall, D-R}.
However, the possible role of corrections to gravity sector is still to be investigated. The importance
of additional terms in gravity sector comes from modifications of Kasner solution near initial singularity
and the fact that Kasner regimes are "basic blocks" of Mixmaster behavior. 

The goal of the present paper is to study  power-law solutions in vacuum Gauss-Bonnet gravity, which 
replace Kasner solution near initial singularity. We chose  the corrections to Einstein-Hilbert action
in the form of Gauss-Bonnet term for two reasons. First, Gauss-Bonnet contribution appears in the string
gravity corrections. Second, this term is the next to Einstein term in Lovelock gravity \cite{Lovelock}.
In the latter theory the equations of motion are the second order derivative equations for the metrics
coefficients in all levels, and the Lovelock gravity entered
recently in a new stage of intense investigations, mainly in the area of black hole solutions \cite{Cha,Ma}
and related thermodynamical properties \cite{Paddy}.
It should be noted that in 4+1 and 5+1 worlds, the action consisting of Einstein and Gauss-Bonnet term is the exact action
of Lovelock gravity. Since in 4+1 case the pure Gauss-Bonnet gravity is a analog of 2+1 dimensional Einstein
gravity (a theory with no nontrivial vacuum solutions \cite{three}), 
we restrict ourself by the case of 5+1 dimensions.

Apart from the Gauss-Bonnet combination, other terms, quadratic in curvature 
lead to higher-derivative equations of motion, and this can modify cosmology
of the early Universe in a very dramatic way, including loss of stability of several important classical solutions
(see, for example \cite{alan} for isotropic and \cite{hervik, we} for anisotropic Universe). It is interesting that
next to second order corrections in string gravity also lead to higher-derivative equations with corresponding
modifications of cosmological solutions \cite{BentoB, Sami}. In $3+1$ dimensions the Gauss-Bonnet term does not 
contribute to equations of motion, though terms of the form $F(GB)$, where $F$ is some function, do contribute
and give rise to some interesting cosmological scenarios (for recent progress see, for example, 
\cite{Odintsov,Odintsov1}),
however, also resulting in higher-derivative equations. We do not study theories with such kind of equations of
motion in the present paper.

There exist several attempts to study anisotropic multidimensional cosmological solutions in 
Gauss-Bonnet gravity. However, as the main motivations for these studies was to find a conditions for
dynamical reduction (details of this approach can be found, for example, in \cite{Chodos, Szydl1, Szydl2})
, the choice of metrics have been often designed for this problem. The most common choice is
an anisotropic Bianchi I metrics with 3 equal large and others equal small dimension 
\cite{Demaret, Tomsk, Andrew},
though  other possibilities
(3 group for 3 equal scale factors in 9+1 dimensional Universe \cite{Halpern},  3 different large and other
equal small dimensions \cite{Dabr2}, some other choices \cite{Iv1, Iv2}) have also been investigated in 
multidimensional anisotropic cosmology. In the present paper
we deal with a generalized Bianchi I multidimensional metrics with different scale factors, and do not assume any
additional relations between them.

We consider a 5+1 dimensional theory  with the action
\begin{equation}
S=\int \sqrt{-\mathrm{g}} (R + \alpha GB)\,d^6x,
 \label{1a}
\end{equation}

where $GB$ is the Gauss-Bonnet term
$$GB= R^{iklm}R_{iklm}-4R^{ik}R_{ik}+R^2.$$

We study 6-dimensional generalization of Bianchi I metric 

\begin{equation}
\mathrm{g}_{ik}=diag(-1,a(t)^2,b(t)^2,c(t)^2,d(t)^2,f(t)^2).
 \label{2a}
\end{equation}

Defining the Hubble parameters
$H_{a,b,c,d,e}=\frac{\dot a,\dot b, \dot c,\dot d, \dot f}{a,b,c,d,f}$
it is possible to write equations of motion in the form of first
integral

\begin{equation}
\begin{array}{l}
2H_aH_b+2H_aH_c+2H_aH_d+2H_aH_f+2H_bH_c+2H_bH_d+2H_bH_f+2H_cH_d+2H_cH_f+2H_dH_f\\
\\
+24\alpha\left[
H_aH_bH_cH_d+H_aH_bH_cH_f+H_aH_bH_dH_f+H_aH_cH_dH_f+H_bH_cH_dH_f
\right ] =0,
 \end{array}
 \label{3}
\end{equation}
and five dynamical equations. The first equation of motion has the form

\begin{equation}
\begin{array}{l}
2(\dot H_b+H_b^2)+2(\dot H_c+H_c^2)+2(\dot
H_d+H_d^2)+2(\dot H_f+H_f^2)\\
\\+2H_bH_c+2H_bH_d+2H_bH_f+2H_cH_d+2H_cH_f+2H_dH_f +\\
\\+ 8\alpha [3H_bH_cH_dH_f+ (\dot H_b+H_b^2)(H_cH_d+H_cH_f+H_dH_f) +(\dot H_c+H_c^2)(H_bH_d+H_bH_f+H_dH_f)\\
\\ +(\dot H_d+H_d^2)(H_bH_c+H_bH_f+H_cH_f)+(\dot
H_f+H_f^2)(H_bH_c+H_bH_d+H_cH_d) ] =0,
\end{array}
 \label{4}
\end{equation}
\\
four other equations can be obtained by cyclic transmutation of indexes.

Our goal is to study generalizations of Kasner solution for gravity with
the Gauss-Bonnet term. For a power-law behavior the interval has the form
$ds^2=-dt^2+\sum
t^{2p_i}dx_i^2$, and we have $H_i=p_i/t$, $\dot H_i =
-p_i/t^2$. Substituting these relations into equations of motion, we get the following equations for $p_i$:

\begin{equation}
\begin{array}{l}
t^2(p_1p_2+p_1p_3+p_1p_4+p_1p_5+p_2p_3+p_2p_4+p_2p_5+p_3p_4+p_3p_5+p_4p_5)\\
\\+12\alpha
(p_1p_2p_3p_4+p_1p_2p_3p_5+p_1p_2p_4p_5+p_1p_3p_4p_5+p_2p_3p_4p_5)=0,
 \label{8}
 \end{array}
\end{equation}

\begin{equation}
\begin{array}{l}
t^2[p_2(p_2-1)+p_3(p_3-1)+p_4(p_4-1)+p_5(p_5-1)+p_2p_3+p_2p_4+p_2p_5+p_3p_4+p_3p_5+p_4p_5]\\
\\+4\alpha[3p_2p_3p_4p_5+p_2(p_2-1)(p_3p_4+p_3p_5+p_4p_5)+p_3(p_3-1)(p_2p_4+p_2p_5+p_4p_5)\\
\\+p_4(p_4-1)(p_2p_3+p_2p_5+p_3p_5)+p_5(p_5-1)(p_2p_3+p_2p_4+p_3p_4)]=0,
 \label{9}
 \end{array}
\end{equation}

\begin{equation}
\begin{array}{l}
t^2[p_1(p_1-1)+p_3(p_3-1)+p_4(p_4-1)+p_5(p_5-1)+p_1p_3+p_1p_4+p_1p_5+p_3p_4+p_3p_5+p_4p_5]\\
\\+4\alpha[3p_1p_3p_4p_5+p_1(p_1-1)(p_3p_4+p_3p_5+p_4p_5)+p_3(p_3-1)(p_1p_4+p_1p_5+p_4p_5)\\
\\+p_4(p_4-1)(p_1p_3+p_1p_5+p_3p_5)+p_5(p_5-1)(p_1p_3+p_1p_4+p_3p_4)]=0,
 \label{10}
 \end{array}
\end{equation}

\begin{equation}
\begin{array}{l}
t^2[p_1(p_1-1)+p_2(p_2-1)+p_4(p_4-1)+p_5(p_5-1)+p_1p_2+p_1p_4+p_1p_5+p_2p_4+p_2p_5+p_4p_5]\\
\\+4\alpha[3p_1p_2p_4p_5+p_1(p_1-1)(p_2p_4+p_2p_5+p_4p_5)+p_2(p_2-1)(p_1p_4+p_1p_5+p_4p_5)\\
\\+p_4(p_4-1)(p_1p_2+p_1p_5+p_2p_5)+p_5(p_5-1)(p_1p_2+p_1p_4+p_2p_4)]=0,
 \label{11}
 \end{array}
\end{equation}

\begin{equation}
\begin{array}{l}
t^2[p_1(p_1-1)+p_2(p_2-1)+p_3(p_3-1)+p_5(p_5-1)+p_1p_2+p_1p_3+p_1p_5+p_2p_3+p_2p_5+p_3p_5]\\
\\+4\alpha[3p_1p_2p_3p_5+p_1(p_1-1)(p_2p_3+p_2p_5+p_3p_5)+p_2(p_2-1)(p_1p_3+p_1p_5+p_3p_5)\\
\\+p_3(p_3-1)(p_1p_2+p_1p_5+p_2p_5)+p_5(p_5-1)(p_1p_2+p_1p_3+p_2p_3)]=0,
 \label{12}
 \end{array}
\end{equation}

\begin{equation}
\begin{array}{l}
t^2[p_1(p_1-1)+p_2(p_2-1)+p_3(p_3-1)+p_4(p_4-1)+p_1p_2+p_1p_3+p_1p_4+p_2p_3+p_2p_4+p_3p_4]\\
\\+4\alpha[3p_1p_2p_3p_4+p_1(p_1-1)(p_2p_3+p_2p_4+p_3p_4)+p_2(p_2-1)(p_1p_3+p_1p_4+p_3p_4)\\
\\+p_3(p_3-1)(p_1p_2+p_1p_4+p_2p_4)+p_4(p_4-1)(p_1p_2+p_1p_3+p_2p_3)]=0,
 \label{13}
 \end{array}
\end{equation}

Last terms of these equations come from Gauss-Bonnet contribution.
Leaving the general case for a future work, we consider only low-energy and
high-energy limits in the present paper.

In the former case we neglect the Gauss-Bonnet term, and after a simple
algebra we get the known 6-dimensional generalization of the 
Kasner conditions 
\begin{equation}
\begin{array}{l}
p_1^2+p_2^2+p_3^2+p_4^2+p_5^2=1,\\
p_1+p_2+p_3+p_4+p_5=1,
\end{array}
 \label{14a}
\end{equation}
These equations define a 3-dimensional "Kasner sphere" (an analog of
Kasner circle existing in the 3+1 dimensional cosmology) in the 5-dimensional
space of $p_i$.

Near a cosmological singularity
$t\rightarrow 0$, and neglecting all terms containing time, we get from (5)
\begin{equation}
\begin{array}{l}
p_1p_2p_3p_4+p_1p_2p_3p_5+p_1p_2p_4p_5+p_1p_3p_4p_5+p_2p_3p_4p_5=0,
\end{array}
 \label{16}
\end{equation}
This equation contains a sum of products of four different $p_i$;
as the total number of different $p$ is equal to 5, there are four possible terms.
The equation (12) can be rewritten in a more compact form as

\begin{equation}
 \sum_{i<j<k<l} p_ip_jp_kp_l=0.
\end{equation}
It is clear that, like in the General Relativity case, at least one of $p_i$ should be negative.

Taking an appropriate linear combination of the constraint and dynamical equations 
we get in the limit $t\rightarrow 0$  the second useful equation:
\begin{equation}
\begin{array}{l}
p_1(p_1-1)(p_2p_3+p_2p_4+p_2p_5+p_3p_4+p_3p_5+p_4p_5)+\\
\\
p_2(p_2-1)(p_1p_3+p_1p_4+p_1p_5+p_3p_4+p_3p_5+p_4p_5)+\\
\\
p_3(p_3-1)(p_1p_2+p_1p_4+p_1p_5+p_2p_4+p_2p_5+p_4p_5)+\\
\\
p_4(p_4-1)(p_1p_2+p_1p_3+p_1p_5+p_2p_3+p_2p_5+p_3p_5)+\\
\\
p_5(p_5-1)(p_1p_2+p_1p_3+p_1p_4+p_2p_3+p_2p_4+p_3p_4)=0.
\end{array}
 \label{17}
\end{equation}

We rewrite the equation (\ref{17}) in the form:
\begin{equation}
\sum_{i} p_i(p_i-1) \sum_{i \ne j,k \atop j<k} p_jp_k =0.
\end{equation}
The equations (13) and (15) determine a 3-dimensional subset in 5-dimensional space of parameters $p_i$,
which replaces the Kasner sphere in the pure Gauss-Bonnet gravity. 

This form of equations can be easily generalized to dimensions bigger than 5+1. Starting
from a Bianchi I metrics and following the procedure described above, it is possible to derive
similar equations for other dimensions. We checked that the equations (13) and (15) remains the same
for the 6+1 and 7+1 cases (the total number of $p_i$ is now 
bigger which leads to a bigger number of individual
terms in the sums, though the structure of the sums does not change). However,
in these cases, if we continue to work in Lovelock gravity, the cubic curvature terms appear
in the action, modifying further cosmological behavior near a singularity.

We can easily find several particular cases, satisfying Eqs. (13) and (15). First of all, the
5-dimensional Milne solution (one of $p_i$ is equal to $1$, others are zero) is still a solution
for the Gauss-Bonnet gravity. Moreover, now nonzero $p_i$ can be arbitrary. Even more
general modification of the Milne solution is given by the combination of indexes $(a, b, 0, 0, 0)$,
where the values of constants $a$ and $b$ are not constrained.

Using equations (13) and (15) it is possible to study the volume expansion rate. In order to do this
we expand brackets in (15). First, consider terms with minus sign. It is easy to see that these terms
form a set of all possible products of three $p$ with different indexes, and each combination
$p_i p_j p_k$ with $i < j < k$ enters the sum three times.

Going further, let us consider a combination $\sum_{i} p_i \sum_{j<k<l} p_jp_kp_l$.
It contains terms of two possible forms. If the index $i$ coincides with one of the indexes from
the right sum, the resulting term has the structure of terms in remaining part of the Eq. (15), and each
term of this part appears in the considered combination only one time. If the index $i$ is different
from indexes of the right sum, the term is a product of four different $p$. Each combination of indexes
appears four times, and using (13) we see that this part of the combination vanishes. 

Combining these results, we get
$$
\sum_{i} p_i \sum_{j<k<l} p_jp_kp_l = \sum_{i} p_i^2 \sum_{i \ne j,k \atop j<k} p_jp_k = 3\sum_{j<k<l} p_jp_kp_l,
$$
and, hence, either
\begin{equation}
\sum_{i} p_i =3,
\end {equation}
or 
\begin{equation}
\sum_{j<k<l} p_jp_kp_l = 0.
\end {equation}

The abovementioned generalizations of 5-dimensional Milne solution belong to the case described
by eq. (17). In this case we have no information on volume expansion rate.
The equation (16), when it is applicable,
 is an analog of the second equation (11) for the Gauss-Bonnet gravity. It means
that the volume of the Universe increases as a third power of the time instead of linear growth in GR.

In the present paper we have found vacuum power-law solution in Gauss-Bonnet gravity for a flat 
Universe. Future work will be devoted to the influence of spatial curvature, and a possible
problem is searching for analogs of Mixmaster behavior in Gauss-Bonnet gravity. 

\section*{Acknowledgments}

This work is supported by RFBR grant 05-02-17450 and
scientific school grant 1157.2006.2 of the Russian Ministry
of Science and Technology.

\end{document}